\documentclass[runningheads]{llncs}
\usepackage[T1]{fontenc}
\usepackage{graphicx}
\usepackage{amsmath}
\usepackage{hyperref}

\begin{document}
\title{Geometric Learning-Based Transformer Network for Estimation of Segmentation Errors}
%
\titlerunning{GL-Based Transformer Network for Estimation of Segmentation Errors}
%
\author{Sneha Sree\inst{1}\and
Mohammad Al Fahim\inst{1}\and
Keerthi Ram\inst{2}\and            
Mohanasankar Sivaprakasam\inst{1,2}
}
\authorrunning{Sneha Sree et al.}
%
\institute{Indian Institute of Technology Madras, India\and
Healthcare Technology Innovation Centre, IIT Madras, India \\
\email{\{snehacumsali,alfahimmohammad\}@gmail.com, keerthi@htic.iitm.ac.in, mohan@ee.iitm.ac.in}
}
\maketitle              
\begin{abstract}
Many segmentation networks have been proposed for 3D volumetric segmentation of tumors and organs at risk. Hospitals and clinical institutions seek to accelerate and minimize specialists' efforts in image segmentation, but in case of errors generated by these networks, clinicians would have to edit the generated segmentation maps manually. 

\textbf{Problem Statement: } Given a 3D volume and its putative segmentation map, we propose an approach to  identify and measure erroneous regions in the segmentation map. Our method can estimate error at any point or node in a 3D mesh generated from a possibly erroneous volumetric segmentation map, serving as a Quality Assurance tool.

\textbf{Method: }We propose a graph neural network-based transformer based on the Nodeformer architecture to measure and classify the segmentation errors at any point. We have evaluated our network on a high-resolution $\mu$CT dataset of the human inner-ear bony labyrinth structure by simulating erroneous 3D segmentation maps. Our network incorporates a convolutional encoder to compute node-centric features from the input $\mu$CT data, the Nodeformer to learn the latent graph embeddings, and a Multi-Layer Perceptron (MLP) to compute and classify the node-wise errors.  

\textbf{Results: }Our network achieves a mean absolute error of $\sim0.042$ over other Graph Neural Networks (GNN)  and an accuracy of $79.53\%$ over other GNNs in estimating and classifying the node-wise errors, respectively. We also put forth vertex-normal prediction as a custom pretext task for pre-training the CNN encoder to improve the network's overall performance. Qualitative analysis shows the efficiency of our network in correctly classifying  errors and reducing misclassifications. 

 \keywords{3D Segmentation error detection, geometric learning}
\end{abstract}
\section{Introduction}
Medical image segmentation is crucial to isolate and analyze specific structures or regions of interest in a medical image to aid in the diagnosis, treatment planning, and monitoring of diseases or conditions. Deep learning models have evolved in accuracy, versatility, and deployment-readiness for automatic segmentation of various organs across diverse medical imaging modalities ~\cite{ronneberger2015unet},~\cite{roth2018deep},~\cite{Chen2019FullyAM}. Still, automated medical image segmentation needs output review, as models are known to be overconfident, although dealing with natural biological variations and diversity in pathological  presentation. 
There is a need for an automated method of predicting and identifying segmentation errors to aid in improving the segmentation maps in erroneous regions.

\textbf{Related Works: }Many recent works have studied the problem of detecting segmentation errors. Kronman et al. \cite{DBLP:journals/cars/KronmanJ16} proposed a geometrical segmentation error detection and correction method in which they detect segmentation errors by casting rays from the interior of the initial segmentation map to its outer surface. Altman et al. \cite{altman2015framework} created an automatic contour quality assurance method that utilizes a knowledge base of historical data. Chen et al. \cite{chen2015automated} proposed supervised geometric attribute distribution models to identify contour errors accurately. The Reverse Classification Accuracy method \cite{robinson2019automated} identifies failed segmentations to predict the CMRI segmentation metrics, achieving a strong correlation with the predicted metrics and visual quality control scores. Alba et al. \cite{alba2018automatic} utilized a random forest classifier with statistical, pattern, and fractal descriptors to detect segmentation contour failures directly without the need for intermediate regression of segmentation accuracy metrics.
Roy et al. \cite{roy2018bayesian} presented an approach that directly incorporates a quality measure or prediction confidence within the segmentation framework. This measure is derived from the same model, eliminating the need for a separate model to evaluate quality. By leveraging model uncertainty, their approach avoids the requirement of training an independent classifier for evaluation, which could introduce additional prediction errors.

\textbf{Graph Neural Networks }(GNN) are deep learning algorithms that can extract features from complex graph structures through message-passing. They are particularly suited for processing three-dimensional data and extracting geometric features to capture and analyze the data structure \cite{wang2019dynamic}. 
Henderson et al. \cite{henderson2022automatic} proposed a quality assurance tool for identifying segmentation errors in 3D organs-at-risk (OAR) segmentations using a geometric learning method by considering the parotid gland. Their study focuses on the parotid gland in head-neck CT scans. 

Inspired by this work \cite{henderson2022automatic}, we propose a novel segmentation error identification network to predict and classify segmentation errors in the inner ear human bony labyrinth using Nodeformer~\cite{wu2022nodeformer}, an advanced Transformer based GNN. We also investigate the effect of pre-training tasks on improving the encoding of node feature vectors for GNNs. 
The key contributions of our work are:
\textbf{(1) }We propose a novel 3D segmentation error estimation network based on graph learning, capable of handling graphs with millions of nodes generated from 3D segmentation maps. \textbf{(2) }We present VertNormPred, a novel pretext task for pre-training the encoder of our network. It involves predicting the node-wise vertex normals to capture the graph's geometric relationships and surface orientations. \textbf{(3) }We quantitatively and qualitatively evaluate our network against other GNN models to estimate and classify node-wise segmentation errors.
\begin{figure}[tb]
    \includegraphics[width=1.0\textwidth]{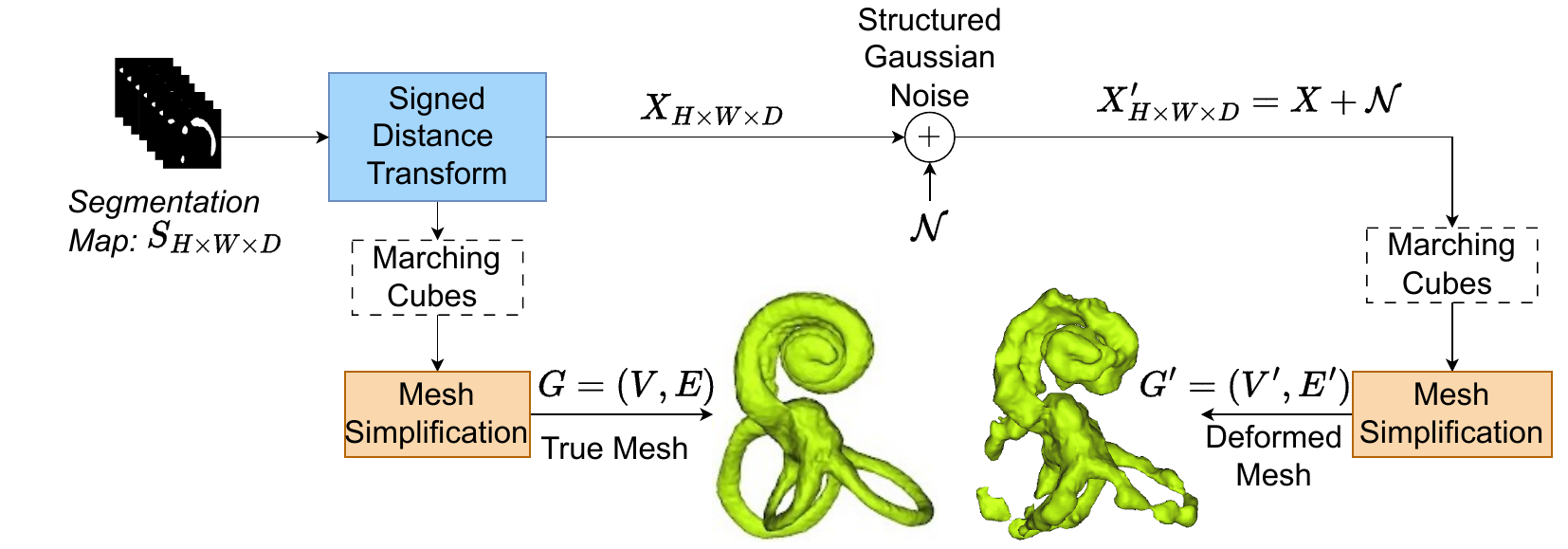}
    \caption{Simulation of true and perturbed meshes for self-supervised learning of segmentation errors.  Each specimen's SDT was perturbed 100 times to produce 100 different deformed segmentations and meshes. The mesh simplification process utilized Taubin smoothing and quadric error decimation techniques to achieve smoother mesh representations.}
    \label{fig1}
\end{figure}

\section{Methods}
\textbf{Formulation: }Let $S$ be the input segmentation map of the $\mu$CT volume $I$. Let $T$ be the true segmentation map of $I$. There exists a deformation on $S$ which operates in the voxel grid to transform $S$ to approach $T$ (limited to nearest neighbors interpolation). The alternative (and finer) domain for mutating $S$ is the surface mesh, computed through a discrete Marching Cubes algorithm~ \cite{10.1145/37402.37422} $f$ on (per-label) extracted contours.

By defining contours as zero-crossings on a signed Euclidean Distance Transform, we have an additional interim domain of the distance transform, which though residing on the voxel grid, offers some unique properties. For instance, take $X'=SDT(S)$ to be the Signed Distance Transform of $S$, and likewise, for the true segmentation map, define $X=SDT(T)$. A dense deformation mapping $S$ to $T$ is modeled conveniently as an additive distortion of $X$ with a structured (sparse) `noise' field: $X'=X+\mathcal{N}$, and recovering $T$ from $S$ becomes estimating and subtracting the noise in $X'$. Further, the discrete distance transform domain can be interpolated to match the resolution of the surface mesh.

Thus, estimating a per-voxel additive correction on $X'$, conditioned on $I$ would lead to determining the location and magnitude of errors in segmentation. This is mapped to learning from ground truth segmentations $T$ through known random perturbations applied in the form $X'=SDT(T)+\mathcal{N}_{sim}$, leading to a self-supervised learning problem, as shown below.

\begin{equation}\label{eq1}
    \begin{split}
    X = SDT(T), \; T = X \leq 0 \\
    X'=X+\mathcal{N}_{sim}, \; S = X'\leq 0
    \end{split}
\end{equation}

Instead of solving this in the SDT domain, we proceed to the mesh domain to setup a per-mesh-vertex estimation of $\mathcal{N}(v)$ conditioned on $I$, which is equivalent to a corrective field in the interpolated SDT space.

\textbf{Graph learning: } The surface mesh of a segmentation map $S$, computed through an operation such as the Discrete Marching Cubes, is representable as a graph $G'=(V', E')$, whose nodes are the mesh vertices, and edges the sides of the triangular faces. 
\begin{equation}\label{eq5}
    G' = (V',E') = f(X') 
\end{equation}
A vertex $v_i$ can be localized in the voxel grid of $I$ to assign an interpolated intensity value. Extending further, a local subvolume in $I$ can be defined around $v_i$. Finally, $v_i$ is connected to nearby vertices forming a local topological arrangement conditioned on image structure. To capture these relationships jointly in the mesh and image domain, we propose to use graph neural networks. 

The learning task is the prediction of node-wise segmentation errors by predicting node-wise Signed Distances (SD) and classifying the node-wise SD into different ranges, given the $\mu$CT subvolume centred at each node $v$, and the entire mesh $G'$. 




The GNN is setup as 
\begin{equation}\label{eq6}
    \mathcal{\hat{N}}\left(v'\right) := h_\theta\left(G^{\prime}, I\right) \quad \forall v' \in V'
\end{equation}
and optimized as
\begin{equation}\label{eq7}
    \theta^* = \arg\min \left\|\mathcal{\hat{N}}-\mathcal{N}_{sim}\right\|_2^2
\end{equation}

\textbf{Modeling: }We propose a graph learning network based on NodeFormer~\cite{wu2022nodeformer}, an advanced Transformer based model designed for efficient node classification on large graphs. NodeFormer incorporates an all-pair message-passing method on adaptive latent structures, enabling information exchange between all nodes by effectively capturing the local and global context. To handle larger graphs, Nodeformer employs the kernelized Gumbel-Softmax operator~\cite{wu2022nodeformer}, enabling scalability to millions of nodes. 

Our intuition behind the model design was, a CNN encoder can capture contextual details from the $\mu$CT data, while the GNN effectively utilizes the local neighborhood of the graph, considering the associated data for each node $v$. By leveraging the graph's local neighborhood based on data, the GNN can analyze the relationships and connectivity between graph elements, allowing the model to incorporate both the image contextual information from $\mu$CT data and the geometric structure of the input. This approach enables the model to exploit the information provided by the local neighborhood of each graph element, enhancing its ability to analyze and process the input data effectively.

\subsection{Architecture}
We choose a CNN consisting of two 3D Conv layers, each followed by ReLU activation functions as a feature extractor to produce node-wise representations of a $5\times5\times5$ $\mu$CT subvolume centered around each node. The extracted node features are embedded with the perturbed graph's edge connectivity information and passed on to the graph transformer network, consisting of three Nodeformer Conv layers. This takes in the graph-embedded node-wise representations and performs all pair message-passing, updating each node’s representation. We consider three Nodeformer Conv layers with eight attention heads, and Batch Normalization and a Leaky ReLU activation function followed each layer.  Finally, a Multi-Layer Perceptron (MLP) consisting of three fully connected layers, wherein each layer was followed by a ReLU activation function, Batch Normalization, and a Dropout regularization, processes the updated node-wise representations to produce node-wise SD predictions (using Tanh activation function in the last layer) or classifications (using Softmax activation function in the last layer).
 \begin{figure}[tbh]
    \centering
     \includegraphics[width=1.0\textwidth]{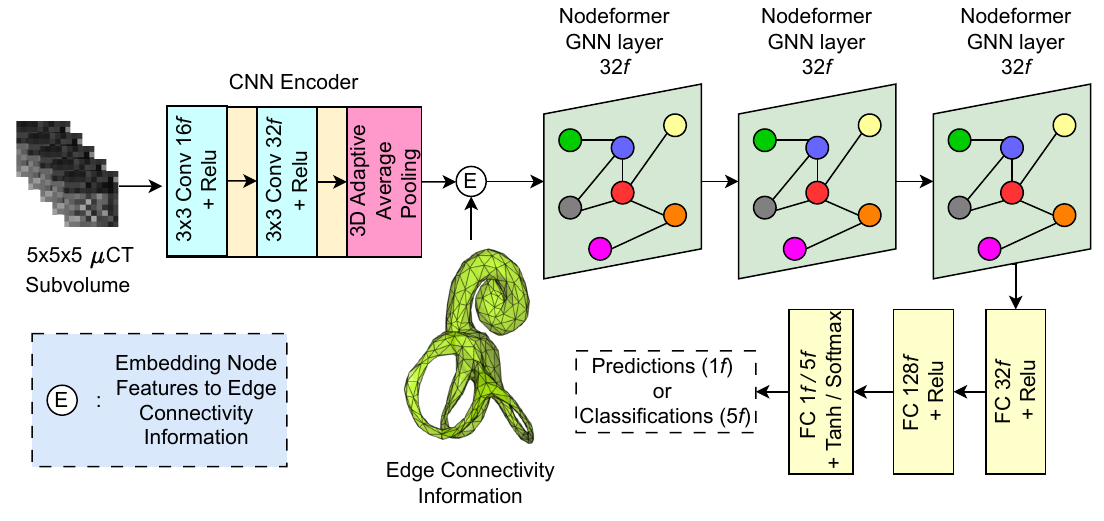}
    \centering
    \caption{The proposed graph learning-based transformer network for predicting and classifying node-wise errors. $\#$f represents the number of output channels/nodes. Given the $5\times5\times5$ $\mu$CT subvolume centred at each node and the edge connectivity information of the perturbed mesh, the model predicts the errors at each node.}
\label{fig2}
\end{figure}

For classification, predicted node-wise SDs are classified into five classes as shown in Fig. \ref{fig5}, ranging from SDs of $-0.16$ mm to $+0.16$ mm. Nodes falling into the higher end of the range, exceeding $+0.16$mm, suggest the occurrence of out-segmentation errors in broad regions. Conversely, nodes with SDs below $-0.16$mm indicate in-segmentation errors specifically within narrow regions. These observations highlight the correlation between SDs and the likelihood of realistic segmentation errors in different regions of interest. Fig. \ref{fig2} illustrates the proposed network architecture for node-wise SDs prediction and classification.

\subsection{Pre-training Tasks}
Towards improving the prediction of node-wise SDs, we incorporated the pre-training transfer learning technique by initializing the model with pre-trained weights obtained from training on different pretext tasks. This approach allows leveraging the knowledge and representations learned during the pretext task to tackle the mainstream tasks~\cite{yosinski2014transferable}.
\begin{figure}[b!]
    \centering
     \includegraphics[width=1.0\textwidth]{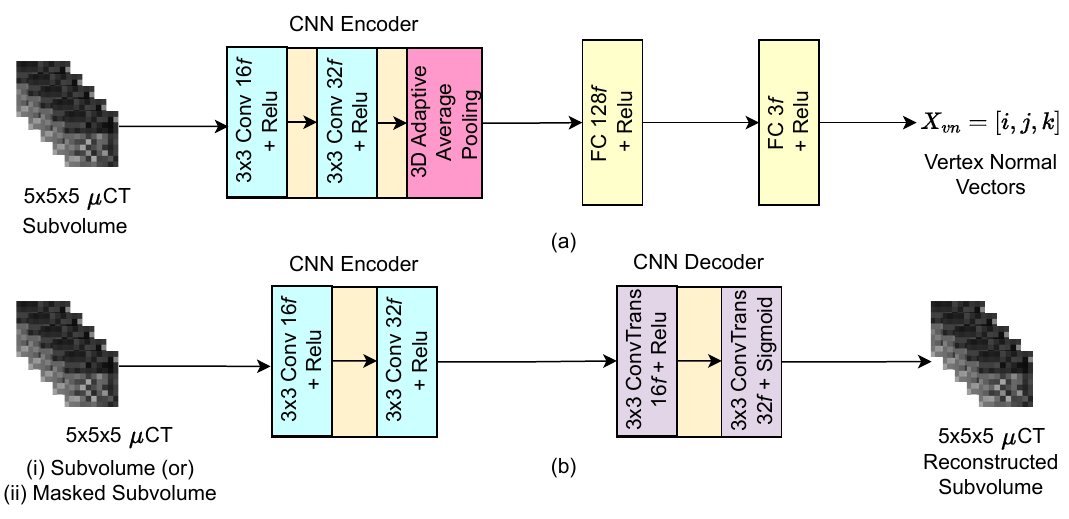}
    \centering
    \caption{a)Vertex Normal Prediction (VertNormPred) network predicts the node-wise vertex normals given the $5\times5\times5$ $\mu$CT subvolume centred at each node. b) (i) CT volume Reconstruction (ReconCT) network and (ii) Masked CT volume Reconstruction (MaskReconCT) network reconstructs the $5\times5\times5$ $\mu$CT subvolume given $\mu$CT or pixel-wise randomly masked $\mu$CT subvolume centred at each node, respectively.}
    \label{fig3}
\end{figure}

We considered the following three pretext tasks: our custom 1) Vertex Normal Prediction (VertNormPred), 2)  $\mu$CT volume Reconstruction (ReconCT), and 3) Masked  $\mu$CT volume Reconstruction (MaskReconCT) tasks.
In the VertNormPred task, we train the CNN model shown in Fig.~\ref{fig3}$(a)$ to predict the node-wise vertex normal $X_{vn}$ given the $5\times5\times5$ $\mu$CT subvolume centred around a node. While generating the dataset using the marching cubes algorithm, we also obtained the ground truth node-wise vertex normals for each mesh. This task enabled the model to capture geometric relationships and surface orientations. Since neighboring nodes and their orientations influence node-wise SDs~\cite{chen2019learning}, understanding surface properties through vertex normal prediction significantly improved the accuracy of the SDs predictions.

In the ReconCT task, we train an encoder-decoder network illustrated in Fig.~\ref{fig3}$(b)$ to reconstruct $5\times5\times5$ $\mu$CT subvolumes. This task allowed the CNN encoder to extract essential features from the node-wise $\mu$CT data.

In the MaskReconCT ~\cite{he2021masked} task, we focus on reconstructing pixel-wise randomly masked $5\times5\times5$ $\mu$CT subvolumes using an encoder-decoder network shown in Fig.~\ref{fig3}$(b)$. We train the model to infer missing regions in the data. By learning to fill these gaps, the model becomes more adept at estimating SDs, especially when parts of the $\mu$CT are incomplete.

We initialized the CNN encoder of our model with the pre-trained weights obtained from the CNN encoder of the models shown in Fig.~\ref{fig3} from these pretext tasks to facilitate node-feature extraction. The pretext tasks: VertNormPred, ReconCT, and MaskReconCT, improved the model in capturing the $\mu$CT bony labyrinth structure for the mainstream task of prediction/classification of node-wise SD.

\section{Dataset Description}
We use the publicly available OpenAIRE's human bony labyrinth dataset \cite{WIMMER2019104782} to evaluate the method. The dataset consists of clinical Computed Tomography (CT) volumes, co-registered high-resolution micro-CT ($\mu$CT) volumes,  segmentation maps, and surface models of $23$ human bony labyrinths. We used $22$ specimens of $\mu$CT volumes and their corresponding segmentation maps.

\subsection{Generation of Training Data}
\begin{figure}[h]
     \centering
     \includegraphics[width=1.0\textwidth]{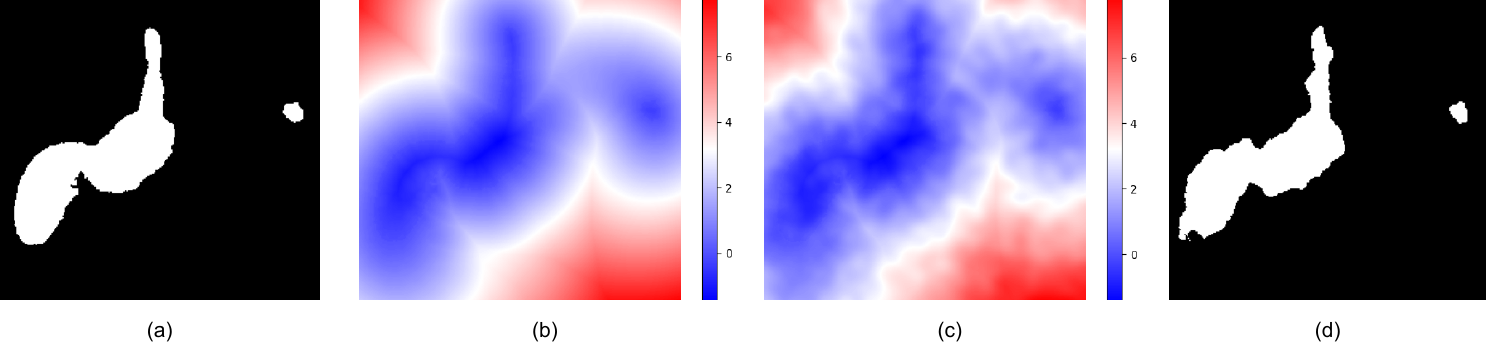}
     \caption{One of the slices of (a) true segmentation map, (b) distance transform, (c) perturbed distance transform after addition of noise to distance transform and (d) perturbed segmentation map obtained from the perturbed distance transform (c)}
     \label{fig4}
 \end{figure}

\begin{figure}[b!]
    \centering
     \includegraphics[width=1.0\textwidth]{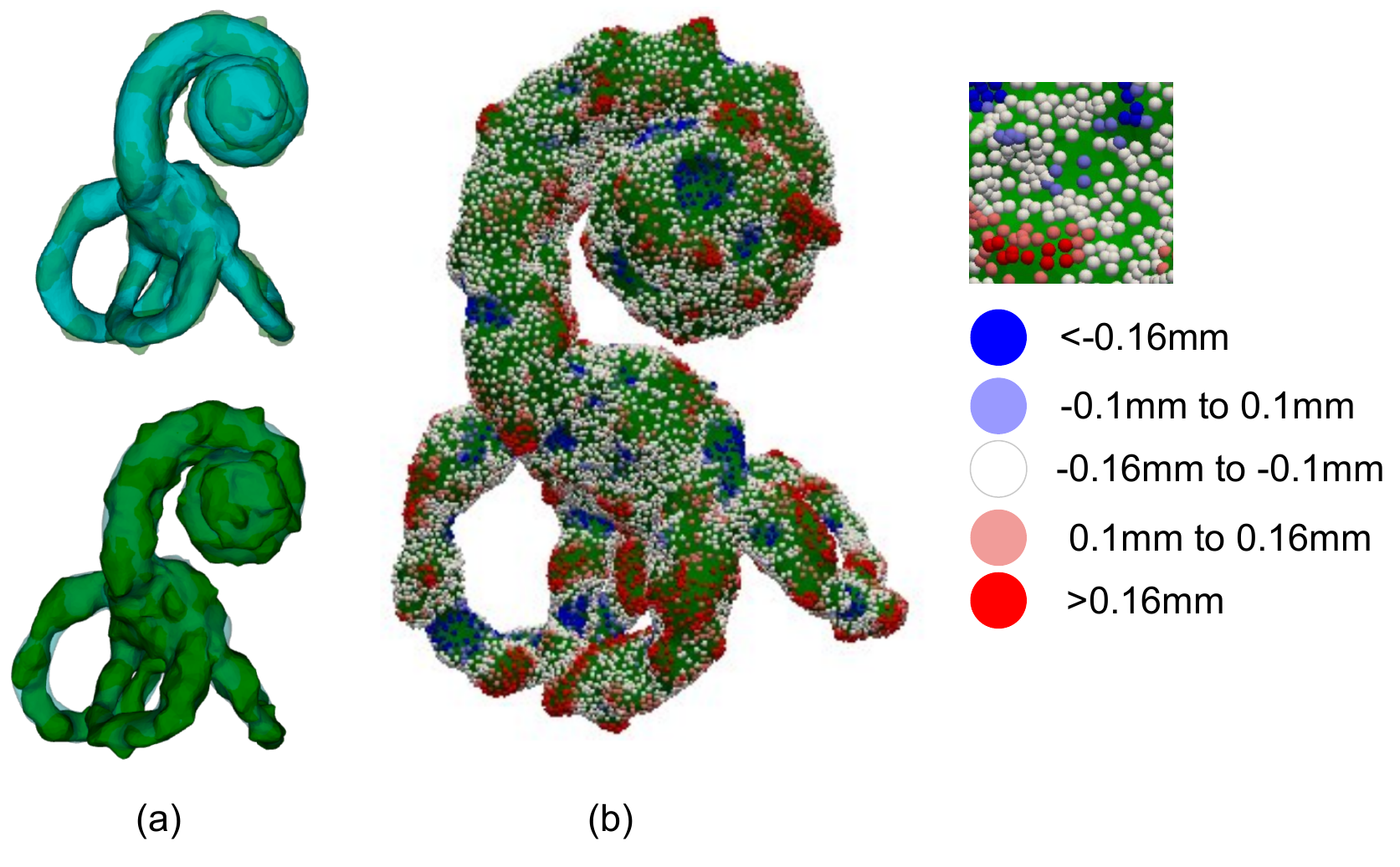}
    \centering
    \caption{Visualization of the true, perturbed meshes, and the node-wise SD classes. (a) At the top, the true mesh (in blue) is overlaid with the perturbed mesh (in green). At the bottom, the perturbed mesh (in green) is overlaid with the true mesh (in blue). The overlapping region between the true and perturbed meshes reveals where internal and external segmentation errors occur. (b) The node-wise SDs in the perturbed mesh are distributed into five classes indicated by colours varying from red to blue, and the class ranges are shown above.}
\label{fig5}
\end{figure}

We generate the perturbed segmentation maps by perturbing the SDT by addition of noise of the true segmentation map 100 times, ensuring the Hausdorff distances of the perturbed segmentation maps are in the range of ($7- 65$). Fig.~\ref{fig4} illustrates the simulation of a perturbed segmentation map obtained from a perturbed SDT. 

We use the marching cubes algorithm to obtain the triangular mesh manifolds of the perturbed segmentation maps. The complex geometry of the human bony labyrinth led to generating a mesh with numerous triangles, resulting in a graph with nodes in the order of $10^5$. We use Taubin smoothing~\cite{inproceedings} and quadratic error decimation techniques to smoothen the mesh. We consider the mesh vertices as nodes ($V$) of the graph and the sides of the triangular faces of the mesh as edges ($E$). The simulation of true and deformed mesh is shown in Fig.~\ref{fig1}.

To calculate the node-wise SD, we perform bi-linear interpolation between the nodes of the perturbed mesh and the voxels of the ground truth SDT. Note that the generated node-wise errors correspond to the node-wise SDs of the true segmentation. For classification, we split these node-wise SDs into five classes ranging from -0.16mm to +0.16mm.

\section{Experiments and Results}
Towards fine-grained prediction of node-wise SDs, we trained and evaluated our model for regression of node-wise SDs. To also identify the errors in different ranges, we trained and evaluated our model for classification to classify the predicted node-wise SDs into different classes.

For all the experiments, we considered 1400 perturbed structures for training, 200 for validation, and 600 for testing.  

We have quantitatively and qualitatively evaluated our model against Spline Conv \cite{fey2018splinecnn} and GAT \cite{veličković2018graph} based GNN models for regression and classification of node-wise SDs. We have also evaluated the models using pre-trained weights from the three pretext tasks. We also performed ablation studies to understand the contribution of each block in our proposed model.

\subsection{Implementation details}
We train the network in Fig.~\ref{fig3}$(a)$ for the VertNormPred task, where we minimize the $Cosine\ Similarity\ loss$ between the predicted and ground truth node-wise vertex normals. We train the network in Fig.~\ref{fig3}$(b)$ for ReconCT and MaskReconCT tasks, where we minimize the $L1\ loss$ between the generated and original $5\times5\times5$ $\mu$CT node-wise subvolumes.

For regression of node-wise SDs, we train the models to minimize the $Smooth$ $L1\ loss$ between the predicted node-wise SDs and the node-wise SDs obtained using interpolation (GT SDs). We used the Mean Absolute Error (MAE) and Mean Square Error (MSE) metrics to quantify the performance of the models trained for regression. For the classification of node-wise SDs, we train the models to minimize the $Cross\ Entropy\ loss$ between the predicted and GT SD classes. We used the F1 score, Precision, Recall, and Accuracy metrics to quantify the performance of the models trained for classification.
\begin{table}[tb]
\centering
\caption{Comparison of Nodeformer with different pre-trained weights against other models for regression of node-wise SDs.}
\label{tab1}
\small
\resizebox{0.65\textwidth}{!}{%
\begin{tabular}{cccc}
\hline
\textbf{GNN}        & \textbf{Pretraining}   & \textbf{MAE} $\downarrow$    & \textbf{MSE} $\downarrow$    \\ \hline
Spline              & ReconCT               & 0.06994          & 0.00986          \\
GAT                 & -                     & 0.06946          & 0.00913          \\
GAT                 & VertNormPred          & 0.0694           & 0.00968          \\
Spline              & MaskReconCT           & 0.06783          & 0.00802          \\
GAT                 & ReconCT               & 0.06755          & 0.00884          \\
GAT                 & MaskReconCT           & 0.06705          & 0.00903          \\
Spline              & -                     & 0.06032          & 0.00762          \\
Spline              & VertNormPred          & 0.05728          & 0.00757          \\
\textbf{Nodeformer} & \textbf{-}            & \textbf{0.04536} & \textbf{0.00475} \\
\textbf{Nodeformer} & \textbf{MaskReconCT}  & \textbf{0.04397} & \textbf{0.00451} \\
\textbf{Nodeformer} & \textbf{ReconCT}      & \textbf{0.04254} & \textbf{0.00444} \\
\textbf{Nodeformer} & \textbf{VertNormPred} & \textbf{0.04182} & \textbf{0.00429} \\ \hline
\end{tabular}%
}
\end{table}
We trained all the networks for 100 epochs, using a learning rate of $1\mathrm{e}^{-3}$ and a cosine annealing scheduler with a weight decay of $1\mathrm{e}^{-3}$. Both the regression and classification models utilized the AdamW optimizer, while the pre-training networks employed the Adadelta optimizer. Models are implemented using PyTorch and PyG~\cite{Fey/Lenssen/2019}, and the training process was carried out in a workstation using an i5-1035G4 CPU and NVIDIA $24GB$ RTX $3090$ GPU.

\subsection{Results and Discussion}
In Table~\ref{tab1}, it can be observed that our model built upon Nodeformer can predict node-wise SDs efficiently, and additionally, using pre-trained weights improved the prediction. Among the evaluated models, our model with a CNN encoder initialized with VertNormPred pre-trained weights yielded the lowest MAE score of $0.04182$. This signifies a substantial improvement of $\sim30.6$\%\ compared to Spline Conv GNN without any pre-trained weights.

\begin{table}[h]
\centering
\caption{Comparison of Nodeformer with different pre-trained weights against other models for classification of node-wise SD classes.}
\label{tab2}
\small
\resizebox{1.0\textwidth}{!}{%
\begin{tabular}{cccccc}
\hline
\textbf{GNN}        & \textbf{Pretraining}  & \textbf{f1 Score} $\uparrow$& \textbf{Precision} $\uparrow$ & \textbf{Recall} $\uparrow$ & \textbf{Accuracy(\%)} $\uparrow$\\ 
\hline
Spline              & MaskReconCT           & 0.4872            & 0.5445             & 0.5124          & 66.3                  \\
GAT                 & VertNormPred          & 0.5186            & 0.605              & 0.5398          & 69.03                 \\
GAT                 & MaskReconCT           & 0.5024            & 0.5649             & 0.5425          & 71.22                 \\
Spline              & VertNormPred          & 0.5367            & 0.612              & 0.5487          & 71.76                 \\
GAT                 & ReconCT               & 0.5746            & 0.6289             & 0.5927          & 72.17                 \\
Spline              & ReconCT               & 0.5871            & 0.6136             & 0.614           & 72.28                 \\
GAT                 & -                     & 0.5623            & 0.6487             & 0.567           & 72.4                  \\
Spline              & -                     & 0.5582            & 0.6181             & 0.5779          & 71.53                 \\
\textbf{Nodeformer} & \textbf{MaskReconCT}  & \textbf{0.5986}   & \textbf{0.6693}    & \textbf{0.589}  & \textbf{74.55}        \\
\textbf{Nodeformer} & \textbf{ReconCT}      & \textbf{0.6695}   & \textbf{0.72}      & \textbf{0.7131} & \textbf{76.57}        \\
\textbf{Nodeformer} & \textbf{-}            & \textbf{0.6899}   & \textbf{0.7343}    & \textbf{0.6693} & \textbf{78.82}        \\
\textbf{Nodeformer} & \textbf{VertNormPred} & \textbf{0.6943}   & \textbf{0.7384}    & \textbf{0.6835} & \textbf{79.53}       \\ \hline
\end{tabular}%
}
\end{table}

In Table~\ref{tab2}, our model, initialized with pre-trained encoder weights of the VertNormPred task, gave an overall accuracy of $79.53\%$. This signifies a substantial improvement of $\sim8\%$ in accuracy compared to the Spline Conv GNN without any pre-training task, indicating a significant improvement in the model's ability to identify different ranges of segmentation errors. 

Tables \ref{tab1} and \ref{tab2} show that our model has benefited from using the pre-trained weights of the VertNormPred task, indicating that the prediction of the node-wise vertex normals during pre-training has helped the encoder of our model in capturing the intricate surface orientations and geometric inter-node relationships in the bony labyrinth structure. This has helped further improve the prediction of node-wise SDs.

\begin{figure}[h]
    \centering
     \includegraphics[width=1.0\textwidth]{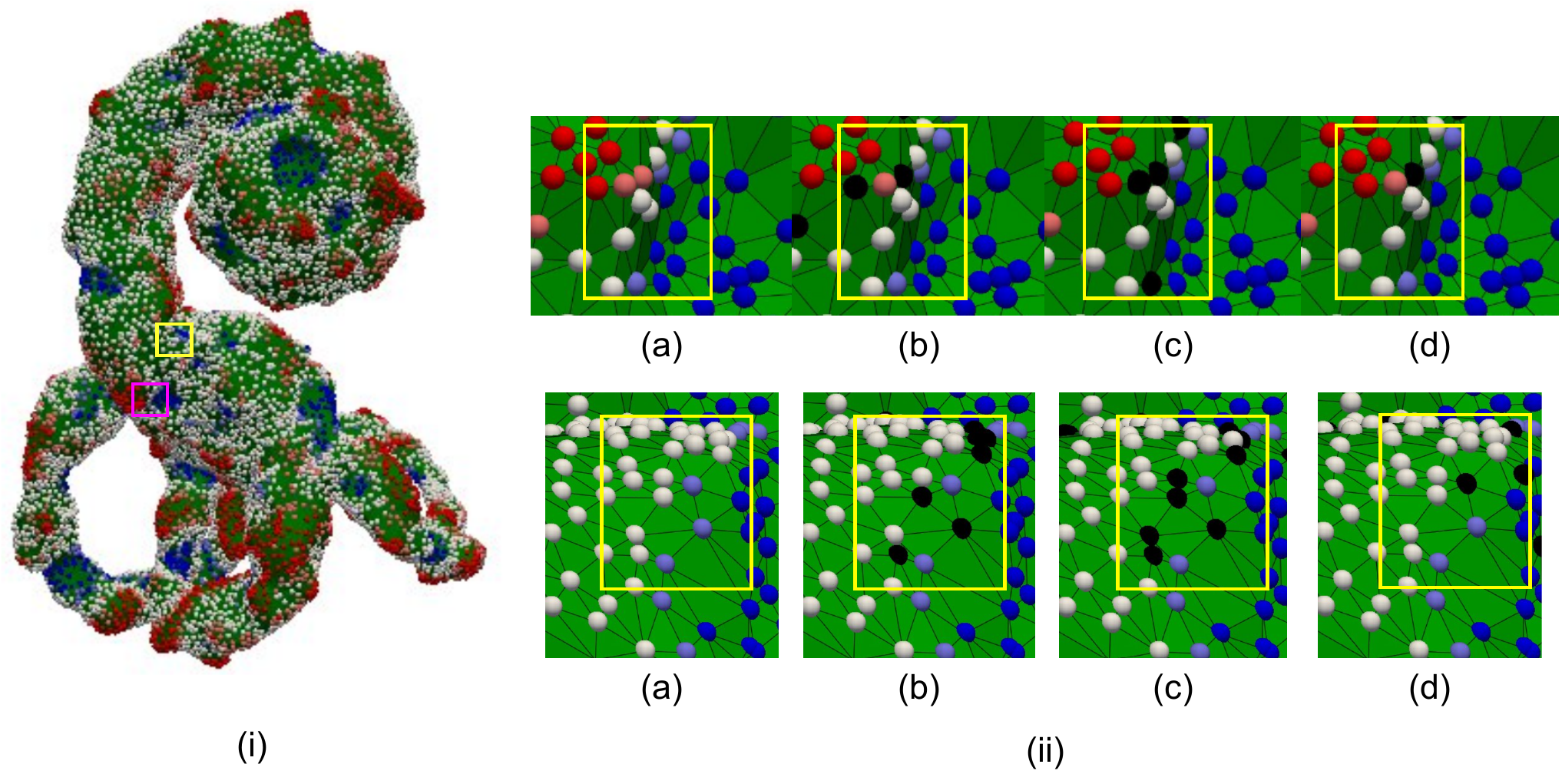}
    \centering
    \caption{Visual illustration of the classification of node-wise SDs by all the models. (i) The actual node-wise SD classes in the perturbed mesh. (ii) Two distinct regions within the graph to showcase the predicted node classes compared to the ground truth node classes. The first and second rows show the zoomed-in regions in the pink and yellow boxes in the perturbed mesh (i). (a) GT perturbed node-wise SD classes, (b)-(d) predicted node-wise SD classes by Spline Conv, GAT, and our model, respectively. The black-coloured nodes denote incorrect predictions. The yellow boxes in (ii) (a)-(d) show how well our model can classify the node-wise SD classes with respect to the GT node-wise SD classes with the least number of black nodes.}
\label{fig6}
\end{figure}

 \begin{figure}[h!]
    \centering
    \includegraphics[width=1.0\textwidth]{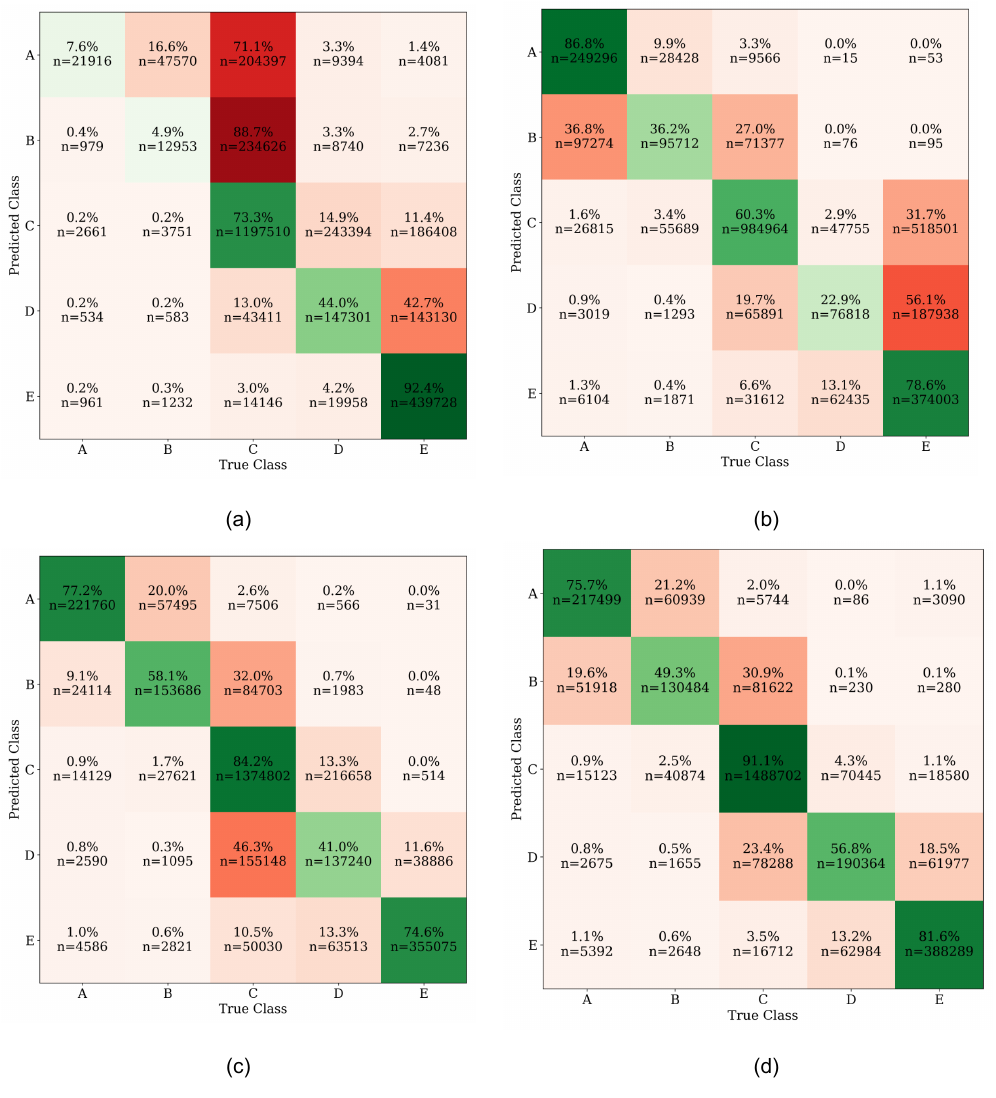}
    \centering
    \caption{Confusion Matrices of (a) CNN-MLP: Node features from the CNN encoder are directly given to the MLP for classification, (b) GNN-MLP: Node feature vectors are obtained from a linear layer instead of the CNN encoder are passed on to the Nodeformer and MLP for classification, (c) our complete model, and (d) our model with the CNN encoder initialized with the VertNormPred pre-trained weights. Labels A: (<-0.16mm), B: (-0.16mm to -0.1mm), C: (-0.1mm to 0.1mm), D: (0.1mm to 0.16mm), E: (>0.16mm)}
\label{fig7}
\end{figure}

From Fig.~\ref{fig6}, it is evident that our model, using Nodeformer, outperforms the other models in the classification of node-wise SDs. Our model qualitatively exhibits improved classification of node-wise SD classes, with significantly fewer black-colored nodes representing incorrect predictions than those obtained using Spline Conv and GAT models. This highlights our model's superior performance and effectiveness in accurately classifying the node classes in the given graph.

In our experiments, we observed that incorporating pre-trained weights from the pretext tasks positively impacted the performance of the models in the regression of node-wise SDs. However, using pre-trained weights for the classification task did not result in much significant improvement. 

\subsection{Ablation Study}
\label{ablative-study}
To evaluate the extent to which Nodeformer effectively learns meaningful information from the geometric structure of the segmentation, we performed an ablation study for the classification task that involved removing the GNN component entirely and directly passing on the node-wise representations from the encoder to the MLP decoder. Also, to evaluate the importance of node feature extraction using the CNN encoder and pre-trained weights, we experimented by passing the $\mu$CT subvolumes through a linear layer as node feature representations to the Nodeformer instead of passing them through the CNN encoder. 

\begin{figure}[t]
    \centering
     \includegraphics[width=1.0\textwidth]{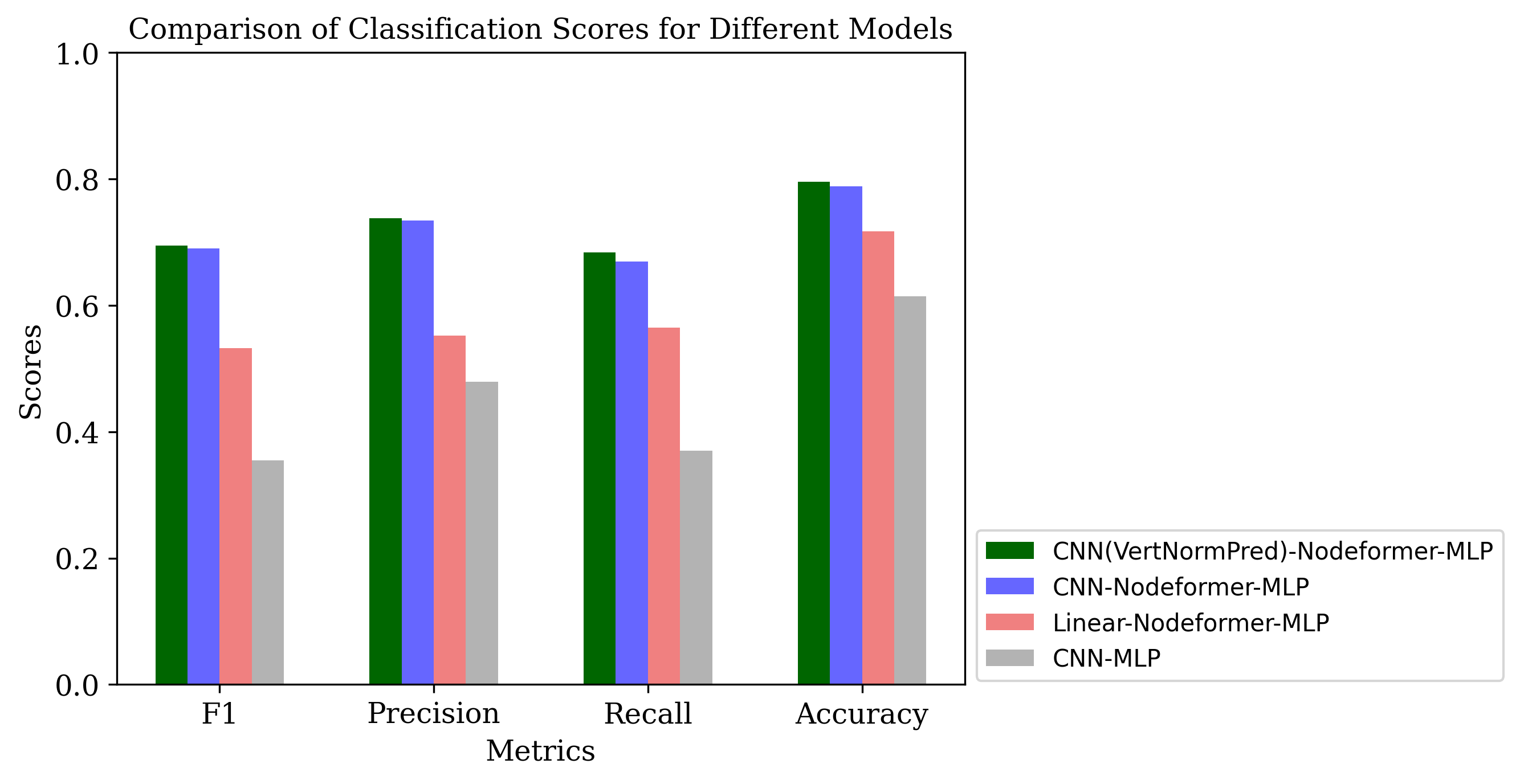}
    \centering
    \caption{ Comparison of Classification Scores between the different blocks of our model, described in section \ref{ablative-study}. The plot provides a visual representation of the distribution and relative performance of the models based on their classification scores.}
\label{fig8}
\end{figure}

Fig.~\ref{fig8} demonstrates the significance of incorporating geometrical structure learning using Nodeformer and the CNN encoder to extract node features in identifying segmentation errors by comparing their performance in classification. Upon removing Nodeformer from our model (CNN-MLP), the classification performance for error identification is notably poor. This emphasizes the importance of Nodeformer in capturing the geometrical information required for error analysis. 

Furthermore, using a linear layer to extract node features from $\mu$CT subvolumes instead of the CNN encoder also resulted in poor performance, as shown in Fig.~\ref{fig8}. This highlights the importance of Conv layers in effectively capturing the node-centred $\mu$CT information necessary for accurate error classification.

Regarding using pre-trained weights, our model with the CNN encoder initialized with pre-trained weights from the VertNormPred task gave the best classification performance regarding accuracy, precision, recall, and F1 score, as shown in Figs. \ref{fig7} and \ref{fig8}. 

By comparing Fig.~\ref{fig7}(a) and Figs.~\ref{fig7}(b)-(d), it can be observed that removing the GNN component (Nodeformer) led to a notable decrease in the model's performance in classifying errors in different ranges. Specifically, it fails to identify internal errors (recall score of 7.6\%). Fig.~\ref{fig7}(b) shows that the Linear-Nodeformer-MLP (GNN-MLP) model can identify internal and external errors but fail to identify the intermediate ones. From Fig.~\ref{fig7}(c) and (d), it is clear that the models using Nodeformer for geometrical structure learning with a CNN encoder for node feature extraction were capable of identifying errors in all ranges and using pre-trained weights reduced misclassification in some classes and significantly improved the recall score.

\section{Conclusion}
Our work introduced a Nodeformer-based graph learning network as a Quality Assurance (QA) tool to evaluate errors in the automatic segmentation of medical images. To our knowledge, this is the first work that addresses segmentation errors in the 3D data of the human inner-ear bony labyrinth structure. The complexity of the inner-ear human bony labyrinth structure gave rise to graphs with nodes in the order of $10^5$. Our network, built upon Nodeformer, can scale up to millions of nodes and easily handle human inner-ear bony labyrinth graphs. To boost the performance of our network, we also proposed a custom Vertex Normal Prediction pretext task for pre-training the CNN encoder of our network. We have evaluated our network against other GNN models with pre-trained weights from different pretext tasks for regression and classification of node-wise segmentation errors. We have qualitatively shown how well our model can correctly classify segmentation errors and reduce misclassifications. We have also conducted an ablation study to show the strengths of individual modules of our network, along with loading the pre-trained weights from the Vertex Normal Prediction pretext task, for classification. This study motivates further research into developing and advancing QA techniques and tools for measuring, classifying, and correcting segmentation errors.

\bibliographystyle{splncs04}
\bibliography{root.bib}


\end{document}